\documentclass[twocolumn,amsmath,footinbib,amssymb,aps,prl,superscriptaddress]{revtex4-1}
\usepackage[latin9]{inputenc}
\usepackage{mathtools}
\usepackage{amsmath}
\usepackage{graphicx}
\usepackage[unicode=true,
 bookmarks=false,
 breaklinks=false,pdfborder={0 0 1},backref=false,colorlinks=false]
 {hyperref}
\hypersetup{colorlinks,linkcolor=blue,citecolor=blue,urlcolor=blue}

\makeatletter

\usepackage{mathrsfs}
\usepackage{wasysym}\usepackage[mathscr]{eucal}
\usepackage{xfrac}
\usepackage{dcolumn}
\usepackage{bm}
\usepackage{color}
\usepackage{tabularx}
\usepackage[normalem]{ulem}
\usepackage[all]{hypcap}
\usepackage[dvipsnames,usenames,table]{xcolor}

\newcommand{\nocontentsline}[3]{}
\newcommand{\tocless}[2]{\bgroup\let\addcontentsline=\nocontentsline#1{#2}\egroup}

\makeatother

\begin{document}
\title{Charge-$4e$ superconductivity from multi-component nematic pairing: \\ Application to twisted bilayer graphene
}
\author{Rafael M. Fernandes}
\affiliation{School of Physics and Astronomy, University of Minnesota, Minneapolis,
Minnesota 55455, USA}
\author{Liang Fu}
\affiliation{Department of Physics, Massachusetts Institute of Technology, Cambridge,
MA 02139 USA}
\begin{abstract}
We show that unconventional nematic superconductors with multi-component order parameter in lattices with three-fold and
six-fold rotational symmetries support a charge-$4e$ vestigial superconducting phase above $T_c$. 
The charge-$4e$ state, which is a condensate of four-electron bound states that preserve the rotational symmetry of the lattice, is nearly degenerate with a competing vestigial
nematic state, which is non-superconducting and breaks the rotational symmetry. 
This robust result is the consequence of a hidden {\it discrete} symmetry in the Ginzburg-Landau theory, which
permutes quantities in the gauge sector and in the crystalline sector
of the symmetry group. We argue that random strain generally favors
the charge-$4e$ state over the nematic phase, as it acts as a random-mass
to the former but as a random-field to the latter. Thus, we propose
that two-dimensional inhomogeneous systems displaying nematic superconductivity,
such as twisted bilayer graphene, provide a promising platform to
realize the elusive charge-$4e$ superconducting phase. 
\end{abstract}
\date{\today}

\maketitle
\textit{Introduction. }The collective behavior of interacting
electrons in quantum materials can give rise to a plethora of exotic phenomena. 
An interesting example is charge-$4e$ superconductivity \citep{Berg2009,Radzihovsky2009,Babaev2010,Agterberg2011,Moon2012,HongYao2017,Yuxuan_review}, an intriguing macroscopic quantum phenomena which was theoretically proposed but is yet to be observed. 
In contrast to standard charge-$2e$ superconductors characterized
by Cooper pairing, a charge-$4e$ superconductor is formed by the condensation of four-electron bound states. 
While a clear manifestation of this phase would be vortices with half quantum flux, $\frac{1}{2}\frac{hc}{2e}$, many
of its basic properties, such as whether its quasi-particle excitation spectrum is gapless or gapped, 
remain under debate \citep{HongYao2017}.

An interesting question is which systems are promising candidates to
realize charge-$4e$ superconductivity. One strategy
is to consider systems that display two condensates, and search for
a stable state where pairs of Cooper pairs are formed even in the
absence of phase coherence among the Cooper pairs. One widely explored
option is the so-called pair-density wave (PDW) state, in which the
Cooper pairs have a finite center-of-mass momentum \citep{Yuxuan_review}.
An unidirectional PDW is described by two complex gap functions $\Delta_{\pm\mathbf{Q}}$
that have incommensurate ordering vectors $\pm\mathbf{Q}$. Charge-$4e$
superconductivity, described by the composite order parameter $\Delta_{\mathbf{Q}}\Delta_{-\mathbf{Q}}$,
is a secondary order that exists inside the PDW state. It has been
proposed that the PDW state can melt in two stages before reaching
the normal state \citep{Berg2009}, giving rise to an intermediate
state in which there is no PDW order, $\left\langle \Delta_{\pm\mathbf{Q}}\right\rangle =0$,
but there is charge-$4e$ superconducting order, $\left\langle \Delta_{\mathbf{Q}}\Delta_{-\mathbf{Q}}\right\rangle \neq0$.

\begin{figure}
\begin{centering}
\includegraphics[width=1\columnwidth]{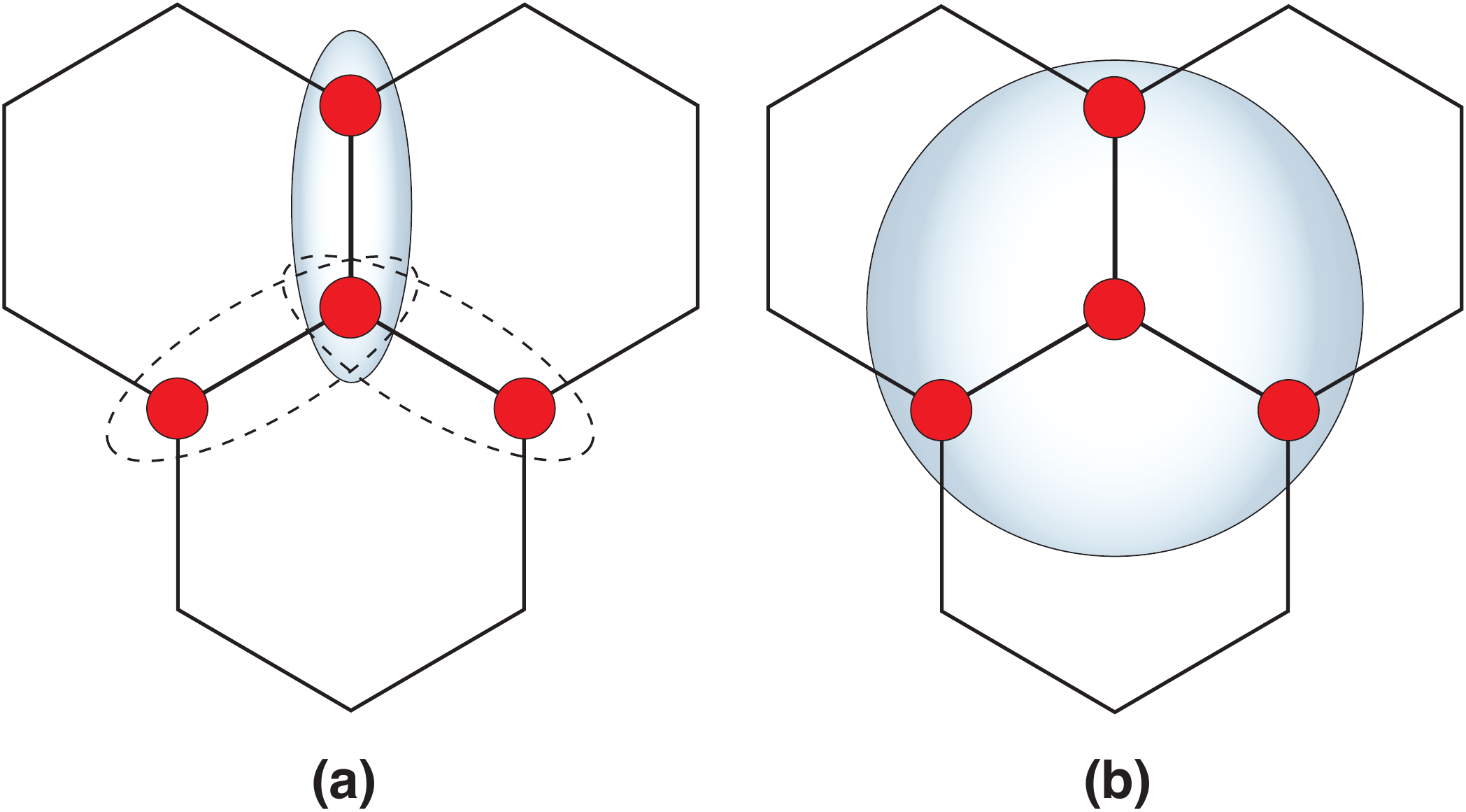}
\par\end{centering}
\caption{A nematic superconducting state in a lattice with three-fold or six-fold
rotational symmetry (here, a honeycomb lattice is shown) is described
by a two-component order parameter $\left(\Delta_{1},\,\Delta_{2}\right)=\Delta_{0}\left(\cos\theta,\,\sin\theta\right)$,
represented here by bound states of electron pairs (red dots). The
ellipses represent, schematically, different orientations $\theta$.
Two competing vestigial phases are supported: (a) a Potts-nematic
phase and (b) a charge-$4e$ phase. In (a), the angle $\theta$ associated
with the nematic director is fixed, breaking the three-fold rotational
symmetry. In (b), the three-fold rotational symmetry is preserved
and a coherent state of bound states of four electrons emerge. In
both (a) and (b), $\left\langle \Delta_{i}\right\rangle =0$, i.e.
charge-$2e$ superconducting order is absent. \label{fig_schematic}}
\end{figure}

Such an intermediate phase is called a vestigial phase
\citep{Nie2014,Tranquada2015,Fernandes_review19}, as it breaks a subset of
the symmetries broken in the primary PDW state. The main drawback
of this interesting idea is the fact that the occurrence of PDW states
in actual materials seems to be rather rare \citep{Yuxuan_review}.
Even from a purely theoretical standpoint, challenges remain in finding
microscopic models that give a PDW ground state rather than a uniform
superconducting ground state. For these reasons, it is desirable to
search for other systems that may host vestigial charge-$4e$ superconductivity.

In this paper, we show that nematic superconductors in hexagonal and
trigonal lattices offer a promising alternative. A nematic superconductor
breaks both the gauge symmetry associated with the phase of the gap
function and the three-fold/six-fold rotational symmetry of the lattice.
Importantly, nematic superconductivity has been experimentally observed
in doped Bi$_{2}$Se$_{3}$ \citep{Bi2Se3_exp1,Bi2Se3_exp2} and in
twisted bilayer graphene \citep{Pablo_nematics}, two systems whose
lattices have three-fold rotational symmetry. Superconducting properties
that do not respect the three-fold lattice symmetry were also observed
in few-layer NbSe$_{2}$, although it is unclear whether this is a
consequence of a nematic pairing state \citep{Hamill2020,Cho2020}.
Unless finite tuning is invoked \citep{Chubukov2019,Wang2020}, 
nematic superconductivity is realized in systems where the order parameter transforms
as a multi-dimensional irreducible representation 
of the relevant point group $G$ \citep{Fu2014,Venderbos_Fu2016,Venderbos18,Su2018,Kozii2019,Pablo_nematics}. Typical examples are  two-dimensional representations  $\left(\Delta_{1},\,\Delta_{2}\right)$ where $\Delta_{1}$ and $\Delta_{2}$ correspond to $p_{x}$-wave/$p_{y}$-wave
gaps or $d_{x^{2}-y^{2}}$-wave/$d_{xy}$-wave gaps. Interestingly,
it has been shown that a secondary composite order parameter $\boldsymbol{\Phi}=\left(\left|\Delta_{1}\right|^{2}-\left|\Delta_{2}\right|^{2},-\Delta_{1}\Delta_{2}^{*}-\Delta_{1}^{*}\Delta_{2}\right)$,
corresponding to Potts-nematic order, can onset even above the superconducting
transition temperature $T_{c}$ \citep{Hecker18,Venderbos18}.

In this paper, we show that the very same mechanism that favors a
vestigial nematic phase also promotes a vestigial charge-$4e$ phase
characterized by a non-zero composite order parameter $\psi=\Delta_{1}^{2}+\Delta_{2}^{2}$
, but $\left\langle \Delta_{i}\right\rangle =0$ (see Fig. \ref{fig_schematic}).
In particular, we find that the effective Ginzburg-Landau theory obtained after integrating
out the normal-state superconducting fluctuations has the same form
for both the nematic order parameter $\boldsymbol{\Phi}$ and
the charge-$4e$ order parameter $\psi$. We show that this is a robust result stemming from the existence of a linear transformation, called
a \textit{perfect shuffle permutation}, that relates $\boldsymbol{\Phi}$
and $\psi$ in the four-dimensional space spanned by $\Delta_{1}$
and $\Delta_{2}$. Such a transformation effectively permutes quantities
in the ``gauge sector'' and in the ``crystalline sector'' of the
group $\mathrm{U}(1)\otimes G$ that defines the symmetry properties
of the system.

This result implies that there are actually two competing vestigial
phases that can onset before long-range superconductivity sets in:
nematic order, as studied previously \citep{Hecker18,Venderbos18},
and charge-$4e$ superconductivity. While higher-order terms in the
superconducting free-energy generally favor the former, we show that the
coupling to random strain can fundamentally alter the balance between
them. This is because random strain acts as a random-field to $\boldsymbol{\Phi}$,
but as a random-mass to $\psi$. Consequently, random strain, intrinsically
present in actual materials, is expected to suppress Potts-nematic
order much more strongly than charge-$4e$ order. We thus conclude
that the most promising candidates to realize vestigial charge-$4e$
superconductivity are relatively inhomogeneous nematic superconductors
with strong superconducting fluctuations, as expected for instance
in quasi-2D systems. This analysis thus suggests that twisted bilayer
graphene \citep{Cao2018a,Cao2018b,Yankowitz19,Efetov19,Sharpe19,STM_Pasupathy19,STM_Andrei19,STM_Perge19,STM_Yazdani19}
offers a potentially viable platform to realize this elusive state
of matter.

\textit{Vestigial nematicity: the standard scenario. } We
consider a nematic superconductor in a lattice with three-fold or
six-fold rotational symmetry, described by a two-component order parameter
$\left(\Delta_{1},\,\Delta_{2}\right)$. For concreteness, hereafter
we will focus on the case where the point group of the lattice is
$D_{6}$, and $\boldsymbol{\Delta}\equiv\left(\Delta_{1},\,\Delta_{2}\right)^{\dagger}$
transforms as the $E_{2}$ irreducible representation (irrep), corresponding
to $\left(d_{x^{2}-y^{2}},\,d_{xy}\right)$-wave gaps. Importantly,
our results are general and hold as long as $\boldsymbol{\Delta}$
transforms as one of the two-dimensional $E$-like irreps of the corresponding
point group $D_{6}$, $D_{3}$, $C_{3v}$, etc. The Ginzburg-Landau superconducting
action expanded to fourth order in $\boldsymbol{\Delta}$ is given by \citep{Sigrist_Ueda,Hecker18,Venderbos18,Chubukov2019}:

\begin{align}
S\left[\boldsymbol{\Delta}\right] & =\int_{q}\Delta_{i,q}^{*}\chi_{ij}^{-1}\left(q\right)\Delta_{j,q}+\frac{u_{0}}{2}\int_{r}\left(\left|\Delta_{1}\right|^{2}+\left|\Delta_{2}\right|^{2}\right)^{2}\nonumber \\
 & +\frac{\gamma}{2}\int_{r}\left|\Delta_{1}\Delta_{2}^{*}-\Delta_{1}^{*}\Delta_{2}\right|^{2}\label{S_Delta}
\end{align}

Here, $\chi_{ij}^{-1}\left(q\right)$ is the inverse superconducting
susceptibility in Fourier space, whereas $u_{0}>0$ and $\gamma$
are Ginzburg-Landau parameters. Furthermore, $q=\left(\mathbf{q},\omega_{n}\right)$
and $r=\left(\mathbf{r},\tau\right)$, where $\mathbf{q}$ is the
momentum, $\omega_{n}$ is the bosonic Matsubara frequency, $\mathbf{r}$
is the position, and $\tau$ is the imaginary time.
Note that $S$ has an enlarged continuous rotational symmetry $\Delta_1 \pm i \Delta_2 \rightarrow e^{\pm i \theta} (\Delta_1 \pm i \Delta_2)$. This emergent continuous rotational symmetry is reduced to discrete ones only when higher-order terms are included, as we discuss later.      

The superconducting ground state depends on $\gamma$: if $\gamma<0$,
the action is minimized by $\boldsymbol{\Delta}=\Delta_{0}\left(1,\,\pm i\right)^{\dagger}$,
corresponding to a time-reversal symmetry-breaking (TRSB) superconductor
that preserves the six-fold rotational symmetry of the lattice. On
the other hand, for $\gamma>0$, the ground state is given by $\boldsymbol{\Delta}=\Delta_{0}\left(\cos\theta,\,\sin\theta\right)^{\dagger}$,
with arbitrary $\theta$. Such a pairing state is called nematic,
as it preserves time-reversal symmetry but lowers the six-fold rotational
symmetry of the lattice to two-fold. It is convenient to construct
the real-valued composite order parameters $\zeta\equiv\boldsymbol{\Delta}^{\dagger}\sigma^{y}\boldsymbol{\Delta}$
and $\boldsymbol{\Phi}\equiv\left(\boldsymbol{\Delta}^{\dagger}\sigma^{z}\boldsymbol{\Delta},\,-\boldsymbol{\Delta}^{\dagger}\sigma^{x}\boldsymbol{\Delta}\right)$,
where $\sigma^{\mu}$ is a Pauli matrix that acts on the two-dimensional
subspace spanned by $\boldsymbol{\Delta}$ \citep{Hecker18,Venderbos18,Fernandes_review19}.
While $\zeta$ transforms as the $A_{2}$ irrep of $D_{6}$, and is
thus related to TRSB, $\boldsymbol{\Phi}$ transforms as the $E_{2}$
irrep, being related to six-fold rotational symmetry breaking. Clearly,
if the ground state is $\boldsymbol{\Delta}=\Delta_{0}\left(1,\,\pm i\right)^{\dagger}$,
$\zeta\neq0$ but $\boldsymbol{\Phi}=0$. On the other hand, if $\boldsymbol{\Delta}=\Delta_{0}\left(\cos\theta,\,\sin\theta\right)^{\dagger}$,
$\zeta=0$ while $\boldsymbol{\Phi}\neq0$. The sign of $\gamma$
is ultimately determined by microscopic considerations. While weak-coupling
calculations tend to favor $\gamma<0$ \citep{Nandkishore2012,Kozii2019,Chubukov2019},
the presence of strong spin-orbit coupling or of density-wave/nematic
fluctuations can tip the balance in favor of the nematic superconducting
state \citep{Fu2014,Venderbos_Fu2016,Kozii2019,Fernandes2013}. Hereafter,
we will assume one of these microscopic mechanisms as the source of
$\gamma>0$.

The nematic superconducting state supports a vestigial nematic phase,
i.e. a phase in which the composite nematic order parameter is non-zero,
$\left\langle \boldsymbol{\Phi}\right\rangle \neq0$, but superconducting
order is absent, $\left\langle \boldsymbol{\Delta}\right\rangle =0$
(see Fig. \ref{fig_schematic}(a)). To see this, we follow the procedure
outlined in Ref. \citep{Fernandes_review19} and first note that the
quartic terms in Eq. (\ref{S_Delta}) can be rewritten in terms of
the TRSB bilinear $\zeta=\boldsymbol{\Delta}^{\dagger}\sigma^{y}\boldsymbol{\Delta}$
and the trivial bilinear $\lambda\equiv\boldsymbol{\Delta}^{\dagger}\sigma^{0}\boldsymbol{\Delta}$
as $S^{(4)}=\frac{u_{0}}{2}\int_{r}\lambda^{2}+\frac{\gamma}{2}\int_{r}\zeta^{2}$.
Here, $\sigma^{0}$ is the identity matrix. Now, the Fierz identity
$\sum_{\mu}\sigma_{ij}^{\mu}\sigma_{kl}^{\mu}=2\delta_{il}\delta_{jk}-\sigma_{ij}^{0}\sigma_{kl}^{0}$
implies a relationship between the bilinears, $\zeta^{2}=\lambda^{2}-\Phi^{2}$.
As a result, the quartic term can be rewritten as $S^{(4)}=\frac{u}{2}\int_{r}\lambda^{2}-\frac{\gamma}{2}\int_{r}\Phi^{2}$,
where $u\equiv u_{0}+\gamma$ and, as defined above, $\boldsymbol{\Phi}=\left(\Phi_{1},\Phi_{2}\right)=\left(\boldsymbol{\Delta}^{\dagger}\sigma^{z}\boldsymbol{\Delta},\,-\boldsymbol{\Delta}^{\dagger}\sigma^{x}\boldsymbol{\Delta}\right)$
is the nematic bilinear. Since $\gamma>0$ by assumption, we can perform
Hubbard-Stratonovich transformations to decouple the quartic terms
and obtain:

\begin{align}
S & \left[\boldsymbol{\Delta},\lambda,\boldsymbol{\Phi}\right]=\int_{r}\frac{\Phi^{2}}{2\gamma}-\int_{r}\frac{\lambda^{2}}{2u}\nonumber \\
 & +\int_{q}\Delta_{i,q}^{*}\left[\chi_{ij}^{-1}\left(q\right)+\lambda\sigma_{ij}^{0}-\Phi_{1}\sigma_{ij}^{z}+\Phi_{2}\sigma_{ij}^{x}\right]\Delta_{j,q}\label{S_phi}
\end{align}

Note that $\boldsymbol{\Phi}$ and $\lambda$ have been promoted to
independent auxiliary fields. Because the action is quadratic in $\Delta_{i}$,
the superconducting fluctuations can be exactly integrated out in
the normal state, yielding an effective action for $\boldsymbol{\Phi}$
and $\lambda$ . Since $\lambda$ does not break any symmetries, it
is always non-zero and simply renormalizes the static superconducting
susceptibility. On the other hand, $\boldsymbol{\Phi}$ is only non-zero
below an onset temperature. A large-$N$ calculation \citep{Fernandes2012},
as performed in Ref. \citep{Hecker18}, indicates that $\left\langle \boldsymbol{\Phi}\right\rangle \neq0$
already above $T_{c}$, implying that vestigial nematic order precedes
the onset of superconductivity (see also the Supplementary Material,
SM). Interestingly, a vestigial nematic phase has been recently observed
in doped Bi$_{2}$Se$_{3}$ \citep{Tamegai2019,Lortz2019}.

\textit{Competition between nematicity and charge-$4e$ superconductivity.
}We now show that there is a hidden symmetry between the two-component
real-valued nematic order parameter $\boldsymbol{\Phi}$ and the complex
bilinear $\psi\equiv\Delta_{1}^{2}+\Delta_{2}^{2}$. The latter breaks
the U(1) gauge symmetry and is precisely the charge-$4e$ order parameter
(see Fig. \ref{fig_schematic}(b)). Importantly, $\psi\neq0$ ($\psi=0$)
inside the nematic (TRSB) superconducting state.

To see the unexpected connection between these two order parameters,
we need to consider, besides the real bilinears discussed above, complex
bilinears formed out of the primary order parameter $\boldsymbol{\Delta}$,
since the latter transforms as the irrep $\Gamma=\mathrm{e}^{im\theta}\otimes E_{2}$
of the group $\mathrm{U}(1)\otimes D_{6}$. Writing the order parameter
explicitly as a four-dimensional vector $\boldsymbol{\eta}\equiv\left(\Delta'_{1},\,\Delta''_{1},\Delta'_{2},\,\Delta''_{2}\right)^{T}$,
where the prime (double prime) denotes the real (imaginary) part,
the bilinears are generally given by $\boldsymbol{\eta}^{T}\left(\sigma^{\mu}\otimes\sigma^{m}\right)\boldsymbol{\eta}$.
Here, the first Pauli matrix (with Greek superscript) in the Kronecker
product $\sigma^{\mu}\otimes\sigma^{m}$ refers to the subspace associated
with the two-dimensional irreducible representation $E_{2}$ of the
point group $D_{6}$ (dubbed the crystalline sector), whereas the
second Pauli matrix (with Latin superscript) refers to the subspace
associated with the U(1) group (dubbed the gauge sector). In this
notation, the components of the nematic bilinear become:

\begin{align}
\Phi_{1} & =\boldsymbol{\eta}^{T}\left(\sigma^{z}\otimes\sigma^{0}\right)\boldsymbol{\eta}\nonumber \\
\Phi_{2} & =-\boldsymbol{\eta}^{T}\left(\sigma^{x}\otimes\sigma^{0}\right)\boldsymbol{\eta}\label{real_bilinear}
\end{align}

The other real bilinears are given by $\zeta=\boldsymbol{\eta}^{T}\left(\sigma^{y}\otimes\sigma^{y}\right)\boldsymbol{\eta}$
and $\lambda=\boldsymbol{\eta}^{T}\left(\sigma^{0}\otimes\sigma^{0}\right)\boldsymbol{\eta}$.
The charge-$4e$ bilinear $\psi\equiv\psi'+i\psi''$, on the other
hand, is:

\begin{align}
\psi' & =\boldsymbol{\eta}^{T}\left(\sigma^{0}\otimes\sigma^{z}\right)\boldsymbol{\eta}\nonumber \\
\psi'' & =\boldsymbol{\eta}^{T}\left(\sigma^{0}\otimes\sigma^{x}\right)\boldsymbol{\eta}\label{complex_bilinear}
\end{align}

The key point is that, although the Kronecker product $\left(M\otimes N\right)$
is non-commutative, in the case where $M$ and $N$ are square matrices
it satisfies the property $\left(M\otimes N\right)=\tilde{P}^{T}\left(N\otimes M\right)\tilde{P}$,
where $\tilde{P}$ is the so-called \emph{perfect shuffle permutation
matrix} \citep{Davio1981}. Here, due to the minus sign in the second
equation of (\ref{real_bilinear}), a slightly modified $2\times2$
matrix $P$ is needed:

\begin{equation}
P=\left(\begin{array}{cccc}
1 & 0 & 0 & 0\\
0 & 0 & -1 & 0\\
0 & -1 & 0 & 0\\
0 & 0 & 0 & 1
\end{array}\right)\label{permutation}
\end{equation}

Physically, $P$ permutes quantities from the crystalline and the
gauge sectors of the four-dimensional space spanned by $\boldsymbol{\eta}$.
Note that $P$ is an orthogonal matrix, $P^{-1}=P^{T}=P$. As a result,
upon performing the unitary transformation $\tilde{\boldsymbol{\eta}}=P\boldsymbol{\eta}$,
we see that while the bilinears $\zeta$ and $\lambda$ remain invariant,
$\left(\Phi_{1},\Phi_{2}\right)\rightarrow\left(\psi',\psi''\right)$,
i.e. the nematic bilinear is mapped onto the charge-$4e$ bilinear.
Consequently, provided that the susceptibility in the quadratic term
of Eq. (\ref{S_Delta}) is invariant under the linear transformation
(\ref{permutation}), the effective action in the normal state has
the same functional form with respect to either $\Phi^{2}$ or $\left|\psi\right|^{2}$.
This is the case if we consider the standard susceptibility expression
$\chi_{ij}^{-1}\left(q\right)=\left(r_{0}+q^{2}\right)\delta_{ij}$,
where $r_{0}\propto T-T_{c,0}$ is a tuning parameter and $T_{c,0}$
is the bare superconducting transition temperature (see the SM).

This is the main result of our paper: for the Ginzburg-Landau action in Eq. (\ref{S_Delta}),
which describes a nematic superconducting ground state in a lattice
with three-fold or six-fold rotational symmetry, an instability towards
a vestigial nematic state at $T_{\mathrm{nem}}$ implies an instability
towards a vestigial charge-$4e$ state at the same temperature $T_{4e}=T_{\mathrm{nem}}$.
This degeneracy between nematicity and charge-$4e$ superconductivity is rooted on
the invariance of the action upon a perfect shuffle that permutes
elements from the crystalline and the gauge sectors.

\textit{Selecting nematic or charge-$4e$ order. }While the competition
between vestigial charge-$4e$ and nematic orders is robust, their
degeneracy is lifted by additional terms in the action not considered
in the analysis above. For instance, additional symmetry-allowed terms
in the susceptibility $\chi(q)$ can favor either the charge-$4e$
state, in the case of a hexagonal lattice, or the nematic state, in
the case of a trigonal lattice (see SM). While here we focus on classical
phase transitions, where the dynamics of $\chi(q)$ is not important,
the situation changes in the case of quantum phase transitions, as
the couplings between the bosonic fields $\boldsymbol{\Phi}$ and
$\psi$ and the electrons are expected to generate different types
of bosonic dynamics.

\begin{figure}

\begin{centering}
\includegraphics[width=0.95\linewidth]{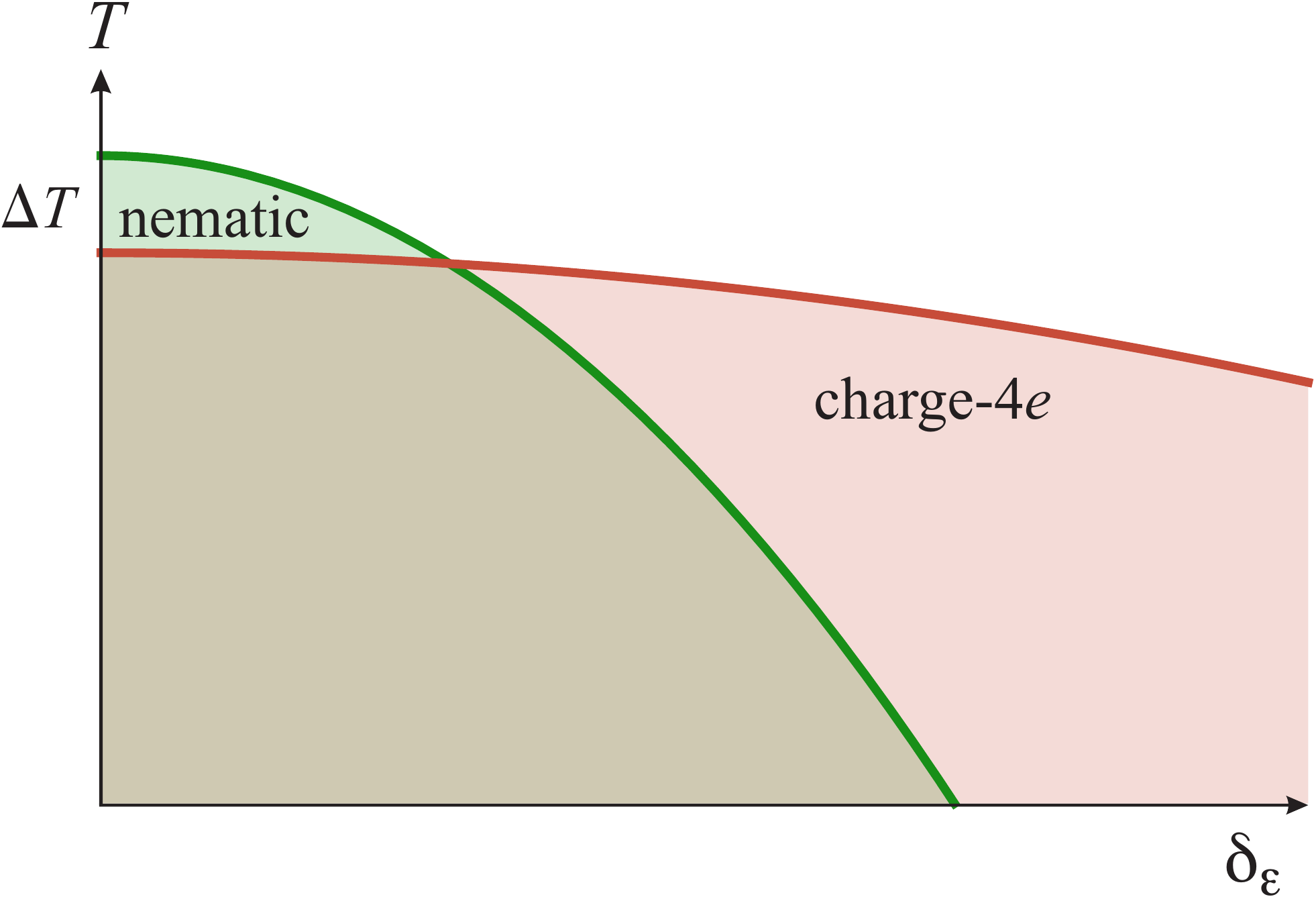}
\par\end{centering}
\caption{Schematic phase diagram of the vestigial nematic (transition temperature
$T_{\mathrm{nem}}$, green) and vestigial charge-$4e$ ($T_{4e}$,
red) phases. Here, $\delta_{\varepsilon}$ represents the strength
of strain inhomogeneity. Because random strain couples as a random-field
to the nematic order parameter but as a random-mass to the charge-$4e$
order parameter, the former is expected to be suppressed much more
strongly than the latter. In the clean system, $\Delta T\equiv T_{\mathrm{nem}}-T_{4e}$
is positive because of the sixth-order term in Eq. (\ref{sixth_order})
that restricts the nematic director to three directions (3-state Potts
nematicity) and lifts the emergent degeneracy between the two vestigial
ordered phases. Note that, as temperature is lowered, a superconducting
transition is expected (not shown here). Whether charge-$4e$ and
nematic orders can coexist in the overlapping region of the phase
diagram remains to be studied. \label{fig_phase_diagram}}

\end{figure}

More importantly, because the nematic order parameter $\boldsymbol{\Phi}$
is real and transforms as the $E_{2}$ irrep of $D_{6}$, there is
a symmetry-allowed cubic term in the nematic action proportional to
$\left(\Phi_{+}^{3}+\Phi_{-}^{3}\right)$, where $\Phi_{\pm}=\Phi_{1}\pm i\Phi_{2}$
\citep{Fu2014, Hecker18,Venderbos18,Cenke_nematics,Fernandes2019}. This term
is related to a particular sixth-order term in the superconducting
action (\ref{S_Delta}) \citep{Sigrist_Ueda}:

\begin{equation}
S^{(6)}\left[\boldsymbol{\Delta}\right]\propto\int_{r}\left(\Delta_{1}+i\Delta_{2}\right)^{3}\left(\Delta_{1}^{*}+i\Delta_{2}^{*}\right)^{3}+\mathrm{h.c.}\label{sixth_order}
\end{equation}

In contrast, because $\psi$ is complex and transforms as the $A_{1}$
irrep of $D_{6}$, such a cubic term is not allowed in the charge-$4e$
action. This cubic term not only favors the nematic order over the
charge-$4e$ order, but it also lowers the symmetry of the nematic
order parameter from U$(1)$ to 3-state Potts \citep{Hecker18,Venderbos18,Cenke_nematics,Li_cold_atoms}.
At first sight, this seems to suggest that it would be challenging
to find a vestigial charge-$4e$ instability occurring before the
onset of vestigial nematic order. While it is possible that charge-$4e$
order could coexist with nematic order and onset at a temperature
between $T_{\mathrm{nem}}$ and $T_{c}$ (the renormalized superconducting
transition temperature), this seems to be a rather contrived scenario.
However, there is an important ingredient missing in the analysis:
the coupling to lattice degrees of freedom. This is particularly important
for nematic order, as it is known to trigger lattice distortions \citep{Fernandes2019}.

We thus introduce the strain tensor $\varepsilon_{ij}=\frac{1}{2}\left(\partial_{i}u_{j}+\partial_{j}u_{i}\right)$,
with $\mathbf{u}$ denoting the lattice displacement vector. Decomposing
it in the irreps of the $D_{6}$ group, there are two relevant modes:
the longitudinal mode, which transforms as $A_{1}$, $\varepsilon_{A}\equiv\varepsilon_{xx}+\varepsilon_{yy}+\varepsilon_{zz}$,
and the shear mode, which transforms as $E_{2}$, $\boldsymbol{\varepsilon}_{E}\equiv\left(\varepsilon_{xx}-\varepsilon_{yy},\,-2\varepsilon_{xy}\right)$.
The leading-order couplings to the nematic and charge-$4e$ orders
are given, respectively, by the linear coupling $\boldsymbol{\varepsilon}_{E}\cdot\boldsymbol{\Phi}$
and by the quadratic coupling $\varepsilon_{A}\left|\psi\right|^{2}$.
While strain can be externally applied, it is intrinsically present
in materials as random strain caused by defects arising in the crystal
growth or device fabrication. The key point is that random strain
acts as a random-field to the Potts-nematic order parameter, but as
a random-mass (also called random-$T_{c}$) to the charge-$4e$ order
parameter.

This distinction is very important, as random-field disorder is known
to be much more detrimental to long-order range order than random-mass
disorder. In the specific case of the 3-state Potts model, random-field
is believed to completely kill the Potts transition in two dimensions,
and to suppress it in three dimensions \citep{RFPM_1984,RFPM_1996,RFPM_2018}.
Thus, one generally expects random strain to tilt the balance between
the competing vestigial charge-$4e$ and nematic orders in favor of
the former. The resulting schematic phase diagram is shown in Fig.
\ref{fig_phase_diagram}.

The condition $T_{4e}>T_{\mathrm{nem}}$ is not enough to ensure a
vestigial charge-$4e$ phase, as one needs to show also that the renormalized
superconducting transition temperature $T_{c}$ inside the charge-$4e$
state is split from $T_{4e}$ \citep{Fernandes_review19}.
A large-$N$ analysis indicates that, for sufficiently anisotropic
quasi-two-dimensional systems, $T_{4e}$ and $T_{c}$ are indeed split \citep{Fernandes2012,Hecker18}.
In this case, while the transition at $T_{4e}$ is XY-like, the transition
at $T_{c}$ is Ising-like due to the coupling $\psi^{*}\left(\Delta_{1}^{2}+\Delta_{2}^{2}\right)$
between the charge-$4e$ and the superconducting order parameters
\citep{HongYao2017}.

\textit{Conclusions. }In this paper, we showed that a nematic superconductor
in lattices with three-fold or six-fold rotational symmetry supports
competing nematic and charge-$4e$ vestigial orders. Such a competition
is rooted on a perfect shuffle permutation that transforms one order
parameter onto the other in the four-dimensional space spanned by
the multi-component superconducting order parameter. We showed that random strain
provides the most promising tuning knob to favor charge-$4e$ superconductivity
over nematic order, due to the fact that it acts as a random-field
disorder to the latter, but as a random-mass disorder to the latter.
These results establish a new class of systems -- nematic superconductors
-- in which charge-$4e$ order may be realized.

Nematic superconductivity has been now observed in doped Bi$_{2}$Se$_{3}$
and in twisted bilayer graphene \citep{Bi2Se3_exp1,Bi2Se3_exp2,Pablo_nematics}.
Based on our results, the most favorable conditions for the observation
of charge-$4e$ superconductivity are systems where superconducting
fluctuations are strong (e.g. quasi-2D superconductors) and where
random-strain is present (e.g. inhomogeneous superconductors). Twisted
bilayer graphene seems to satisfy both conditions, given the ubiquitous
twist angle inhomogeneity \citep{Zeldov19,Pixley2019,Ryu2020,Young2020},
and is thus a promising place to look for this elusive state of matter.
Note that the mechanism proposed here, which relies on an exact discrete symmetry in Ginzburg-Landau theory for multi-component superconductors in general, is different from a recent proposal for
charge-$4e$ superconductivity based on an approximate SU(4) symmetry
of twisted bilayer graphene \citep{Khalaf2020}.
\begin{acknowledgments}
We thank A. Chubukov, P. Orth, J. Schmalian, and J. Venderbos for fruitful discussions. This work 
was supported by the U. S. Department of Energy, Office of Science, Basic Energy Sciences, Materials Sciences and Engineering
Division, under Award No. DE-SC0020045 (R.M.F.) and DE-SC0018945 (L.F.).
\end{acknowledgments}

\bibliographystyle{apsrev}
\bibliography{references}

\clearpage

\begin{widetext}

\begin{center}
\textbf{\large Supplementary material for ``Charge-$4e$ superconductivity from multi-component nematic pairing: Application to twisted bilayer graphene"}
\end{center}

\end{widetext}

\setcounter{equation}{0}
\renewcommand{\theequation}{S\arabic{equation}}

\setcounter{figure}{0}
\renewcommand{\thefigure}{S\arabic{figure}}

\section*{Derivation of the effective action within large-$N$}

Here we derive the explicit form of the effective nematic/charge-$4e$
action by performing a large-$N$ calculation, extending the general
procedure outlined in Refs. \citep{Fernandes2012,Hecker18,Fernandes_review19}.
For a system with $D_{6}$ point group symmetry, the most general
form of the quadratic part of the superconducting action is given
by \citep{Sigrist_Ueda,Venderbos_Fu_Hc2}:

\begin{align}
S^{(2)}\left[\boldsymbol{\Delta}\right] & =\int_{q}\left(r_{0}+q_{\parallel}^{2}+\upsilon q_{z}^{2}\right)\left(\left|\Delta_{1,q}\right|^{2}+\left|\Delta_{2,q}\right|^{2}\right)\nonumber \\
 & +\kappa\int_{q}\left[\left(q_{x}^{2}-q_{y}^{2}\right)\left(\left|\Delta_{1,q}\right|^{2}-\left|\Delta_{2,q}\right|^{2}\right)\right.\nonumber \\
 & \left.+2q_{x}q_{y}\left(\Delta_{1,q}\Delta_{2,q}^{*}+\Delta_{1,q}^{*}\Delta_{2,q}\right)\right]
\end{align}
where $\mathbf{q}_{\parallel}\equiv\left(q_{x},\,q_{y}\right)$ and
$\upsilon>0$, $\left|\kappa\right|<1$ are constants. In terms of
the vector $\boldsymbol{\eta}\equiv\left(\Delta'_{1},\,\Delta''_{1},\Delta'_{2},\,\Delta''_{2}\right)^{T}$,
it can be rewritten as:

\begin{equation}
S^{(2)}\left[\boldsymbol{\eta}\right]=\int_{q}\boldsymbol{\eta}_{q}^{T}\hat{\chi}_{0}^{-1}(q)\boldsymbol{\eta}_{-q}\label{S2}
\end{equation}
with:

\begin{align}
\hat{\chi}_{0}^{-1}\left(q\right) & =\left(r_{0}+q_{\parallel}^{2}+\upsilon q_{z}^{2}\right)\hat{\mathbb{I}}\nonumber \\
 & +\kappa q_{\parallel}^{2}\left(\cos2\theta\,\hat{A}_{1}-\sin2\theta\,\hat{A}_{2}\right)\label{chi_S2}
\end{align}

Here, $\theta\equiv\arctan\left(\frac{q_{y}}{q_{x}}\right)$, the
hat denotes a $4\times4$ matrix and $\hat{\mathbb{I}}\equiv\sigma^{0}\otimes\sigma^{0}$,
$\hat{A}_{1}\equiv\sigma^{z}\otimes\sigma^{0}$, $\hat{A}_{2}\equiv-\sigma^{x}\otimes\sigma^{0}$.
Explicitly:
\begin{align}
\hat{A}_{1} & =\left(\begin{array}{cccc}
1 & 0 & 0 & 0\\
0 & 1 & 0 & 0\\
0 & 0 & -1 & 0\\
0 & 0 & 0 & -1
\end{array}\right)\label{A1}\\
\hat{A}_{2} & =\left(\begin{array}{cccc}
0 & 0 & -1 & 0\\
0 & 0 & 0 & -1\\
-1 & 0 & 0 & 0\\
0 & -1 & 0 & 0
\end{array}\right)\label{A2}
\end{align}

We now move on to the quartic terms. The first one is given by:

\begin{align}
S_{1}^{(4)}\left[\boldsymbol{\Delta}\right] & =\frac{u}{2}\int_{r}\left(\left|\Delta_{1}\right|^{2}+\left|\Delta_{2}\right|^{2}\right)^{2}=\frac{u}{2}\int_{r}\left(\boldsymbol{\eta}^{T}\hat{\mathbb{I}}\boldsymbol{\eta}\right)^{2}\label{aux_S4_1}
\end{align}

Performing a Hubbard-Stratonovich transformation, we introduce the
auxiliary field $\lambda$ and obtain:

\begin{equation}
S_{1}^{(4)}\left[\boldsymbol{\eta},\lambda\right]=-\int_{r}\frac{\lambda^{2}}{2u}+\int_{r}\lambda\left(\boldsymbol{\eta}^{T}\hat{\mathbb{I}}\boldsymbol{\eta}\right)\label{S4_1}
\end{equation}

As for the second quartic term,

\begin{align}
S_{2}^{(4)}\left[\boldsymbol{\Delta}\right] & =-\frac{\gamma}{2}\int_{r}\left[\left(\left|\Delta_{1}\right|^{2}-\left|\Delta_{2}\right|^{2}\right)^{2}\right.\nonumber \\
 & \left.+\left(\Delta_{1}\Delta_{2}^{*}+\Delta_{1}^{*}\Delta_{2}\right)^{2}\right]
\end{align}
there are two different ways to decouple it in terms of auxiliary
fields. In the first case, we introduce the nematic field:

\begin{equation}
S_{2}^{(4)}\left[\boldsymbol{\eta},\boldsymbol{\Phi}\right]=\int_{r}\frac{\Phi^{2}}{2\gamma}-\int_{r}\left[\Phi_{1}\left(\boldsymbol{\eta}^{T}\hat{A}_{1}\boldsymbol{\eta}\right)+\Phi_{2}\left(\boldsymbol{\eta}^{T}\hat{A}_{2}\boldsymbol{\eta}\right)\right]\label{S4_2_nem}
\end{equation}

An alternative way to decouple it is by using the identity:

\begin{align}
\left[\left(\left|\Delta_{1}\right|^{2}-\left|\Delta_{2}\right|^{2}\right)^{2}+\left(\Delta_{1}\Delta_{2}^{*}+\Delta_{1}^{*}\Delta_{2}\right)^{2}\right] & =\nonumber \\
\left[\Delta_{1}^{2}+\Delta_{2}^{2}\right]\left[\left(\Delta_{1}^{*}\right)^{2}+\left(\Delta_{2}^{*}\right)^{2}\right]
\end{align}

We then introduce the charge-$4e$ auxiliary field $\psi$ and obtain:

\begin{equation}
S_{2}^{(4)}\left[\boldsymbol{\eta},\psi\right]=\int_{r}\frac{\left|\psi\right|^{2}}{2\gamma}-\int_{r}\left[\psi'\left(\boldsymbol{\eta}^{T}\hat{B}_{1}\boldsymbol{\eta}\right)+\psi''\left(\boldsymbol{\eta}^{T}\hat{B}_{2}\boldsymbol{\eta}\right)\right]\label{S4_2_charge4e}
\end{equation}
where we defined $\hat{B}_{1}\equiv\sigma^{0}\otimes\sigma^{z}$ and
$\hat{B}_{2}\equiv\sigma^{0}\otimes\sigma^{x}$, i.e.
\begin{align}
\hat{B}_{1} & =\left(\begin{array}{cccc}
1 & 0 & 0 & 0\\
0 & -1 & 0 & 0\\
0 & 0 & 1 & 0\\
0 & 0 & 0 & -1
\end{array}\right)\label{B1}\\
\hat{B}_{2} & =\left(\begin{array}{cccc}
0 & 1 & 0 & 0\\
1 & 0 & 0 & 0\\
0 & 0 & 0 & 1\\
0 & 0 & 1 & 0
\end{array}\right)\label{B2}
\end{align}

The action can thus be written as:

\begin{align}
S_{\mathrm{nem}}\left[\boldsymbol{\eta},\lambda,\boldsymbol{\Phi}\right] & =\int_{q}\boldsymbol{\eta}_{q}^{T}\left[\hat{\chi}^{-1}(q)-\sum_{i}\Phi_{i}\hat{A}_{i}\right]\boldsymbol{\eta}_{-q}\nonumber \\
 & +\int_{r}\frac{\Phi^{2}}{2\gamma}-\int_{r}\frac{\lambda^{2}}{2u}
\end{align}
or, equivalently,

\begin{align}
S_{4e}\left[\boldsymbol{\eta},\lambda,\psi\right] & =\int_{q}\boldsymbol{\eta}_{q}^{T}\left[\hat{\chi}^{-1}(q)-\sum_{i}\psi_{i}\hat{B}_{i}\right]\boldsymbol{\eta}_{-q}\nonumber \\
 & +\int_{r}\frac{\left|\psi\right|^{2}}{2\gamma}-\int_{r}\frac{\lambda^{2}}{2u}
\end{align}

Here, to simplify the notation, we introduced $\psi_{1}=\psi'$ and
$\psi_{2}=\psi''$. Moreover, we defined:

\begin{equation}
\hat{\chi}^{-1}(q)=\hat{\chi}_{0}^{-1}(q)+\lambda\hat{\mathbb{I}}
\end{equation}
which corresponds to shifting the superconducting mass term to $r=r_{0}+\lambda$.
Integrating out the superconducting fluctuations:

\begin{align*}
\int D\boldsymbol{\eta}\exp\left\{ -\int_{q}\boldsymbol{\eta}_{q}^{T}\left[\hat{\chi}^{-1}(q)-\hat{V}\right]\boldsymbol{\eta}_{-q}\right\}  & =\\
\mathcal{N}\exp\left\{ -\frac{1}{2}\int_{q}\mathrm{Tr}\ln\left[\hat{\chi}^{-1}\left(q\right)-\hat{V}\right]\right\} 
\end{align*}
gives the effective actions:\begin{widetext}

\begin{align}
S_{\mathrm{nem}}^{(\mathrm{eff})}\left[\lambda,\boldsymbol{\Phi}\right] & =-\sum_{n=1}^{\infty}\frac{1}{2n}\int_{q}\mathrm{Tr}\left[\hat{\chi}\left(q\right)\hat{V}_{\mathrm{nem}}\right]^{n}+\int_{r}\frac{\Phi^{2}}{2\gamma}-\int_{r}\frac{\lambda^{2}}{2u}\label{S_nem_eff}\\
S_{4e}^{(\mathrm{eff})}\left[\lambda,\psi\right] & =-\sum_{n=1}^{\infty}\frac{1}{2n}\int_{q}\mathrm{Tr}\left[\hat{\chi}\left(q\right)\hat{V}_{4e}\right]^{n}+\int_{r}\frac{\left|\psi\right|^{2}}{2\gamma}-\int_{r}\frac{\lambda^{2}}{2u}\label{S_4e_eff}
\end{align}
\end{widetext}where $\hat{V}_{\mathrm{nem}}=\sum_{i}\Phi_{i}\hat{A}_{i}$
and $\hat{V}_{4e}=\sum_{i}\psi_{i}\hat{B}_{i}$. Now, to relate $\hat{A}_{i}$
and $\hat{B}_{i}$, we note the identity $\left(M\otimes N\right)=\tilde{P}^{T}\left(N\otimes M\right)\tilde{P}$
for $2\times2$ matrices $M,\,N$, where $\tilde{P}$ is the perfect
shuffle permutation matrix \citep{Davio1981}:

\[
\tilde{P}=\left(\begin{array}{cccc}
1 & 0 & 0 & 0\\
0 & 0 & 1 & 0\\
0 & 1 & 0 & 0\\
0 & 0 & 0 & 1
\end{array}\right)
\]
which is orthogonal. In our case, due to the extra minus sign in Eq.
(\ref{A2}), we need a slightly modified orthogonal matrix:

\begin{equation}
P=\left(\begin{array}{cccc}
1 & 0 & 0 & 0\\
0 & 0 & -1 & 0\\
0 & -1 & 0 & 0\\
0 & 0 & 0 & 1
\end{array}\right)\label{permutation}
\end{equation}
which then gives $\hat{B}_{i}=\hat{P}^{T}\hat{A}_{i}\hat{P}$ as defined
in Eqs. (\ref{A1}), (\ref{A2}), (\ref{B1}), and (\ref{B2}). Thus,
we obtain:

\begin{equation}
\mathrm{Tr}\left[\hat{\chi}\left(q\right)\hat{V}_{\mathrm{nem}}\right]^{n}=\mathrm{Tr}\left[\hat{\tilde{\chi}}\left(q\right)\hat{V}_{4e}\right]^{n}
\end{equation}
upon exchanging $\psi_{i}\longleftrightarrow\Phi_{i}$. Here, $\hat{\tilde{\chi}}\equiv\hat{P}^{T}\hat{\chi}\hat{P}$.
Thus, as long as $\kappa=0$ in Eq. (\ref{chi_S2}), we have $\hat{\tilde{\chi}}\equiv\hat{\chi}$,
implying that the two actions -- nematic and charge-$4e$ -- are
identical.

We now proceed to investigate the impact of $\kappa\neq0$. For simplicity,
we focus on the two-dimensional case, setting $\upsilon=0$ in Eq.
(\ref{chi_S2}). We also consider classical finite-temperature phase
transitions. Performing the traces and integrals in Eqs. (\ref{S_nem_eff})-(\ref{S_4e_eff})
and assuming uniform order parameters, we obtain the free-energy densities
(the Ginzburg-Landau constants $\gamma$ and $u$ are rescaled by
a factor of temperature):

\begin{align}
 & F_{\mathrm{nem}}^{(\mathrm{eff})}\left[r,\boldsymbol{\Phi}\right]=\frac{\Phi^{2}}{2}\left\{ \frac{1}{\gamma}-\int_{0}^{\infty}\frac{qdq}{\pi}\,\frac{\left(q^{2}+r\right)^{2}}{\left[\kappa^{2}q^{4}-\left(q^{2}+r\right)^{2}\right]^{2}}\right\} \nonumber \\
 & -\frac{\Phi^{4}}{4}\int_{0}^{\infty}\frac{qdq}{\pi}\,\frac{\left(q^{2}+r\right)^{2}\left[\left(q^{2}+r\right)^{2}+2\kappa^{2}q^{4}\right]}{\left[\kappa^{2}q^{4}-\left(q^{2}+r\right)^{2}\right]^{4}}\nonumber \\
 & -\frac{(r-r_{0})^{2}}{2u}\label{F_nem}
\end{align}
and:

\begin{align}
 & F_{4e}^{(\mathrm{eff})}\left[r,\psi\right]=\frac{\left|\psi\right|^{2}}{2}\left\{ \frac{1}{\gamma}-\int_{0}^{\infty}\frac{qdq}{\pi}\,\frac{\left[\left(q^{2}+r\right)^{2}+\kappa^{2}q^{4}\right]}{\left[\kappa^{2}q^{4}-\left(q^{2}+r\right)^{2}\right]^{2}}\right\} \nonumber \\
 & -\frac{\left|\psi\right|^{4}}{4}\int_{0}^{\infty}\frac{qdq}{\pi}\,\frac{\left[\left(q^{2}+r\right)^{4}+6\kappa^{2}q^{4}\left(q^{2}+r\right)^{2}+\kappa^{4}q^{8}\right]}{\left[\kappa^{2}q^{4}-\left(q^{2}+r\right)^{2}\right]^{4}}\nonumber \\
 & -\frac{(r-r_{0})^{2}}{2u}\label{F_4e}
\end{align}

The mean-field transitions take place when the quadratic coefficients
vanish. This gives the following critical values of $r$, $r_{\mathrm{nem}}=\gamma J_{\mathrm{nem}}$
and $r_{4e}=\gamma J_{4e}$, with:

\begin{align}
J_{\mathrm{nem}} & \equiv\int_{0}^{\infty}\frac{pdp}{\pi}\,\frac{\left(p^{2}+1\right)^{2}}{\left[\kappa^{2}p^{4}-\left(p^{2}+1\right)^{2}\right]^{2}}\\
J_{4e} & \equiv\int_{0}^{\infty}\frac{pdp}{\pi}\,\frac{\left[\left(p^{2}+1\right)^{2}+\kappa^{2}p^{4}\right]}{\left[\kappa^{2}p^{4}-\left(p^{2}+1\right)^{2}\right]^{2}}
\end{align}

An explicit calculation gives:

\begin{align}
J_{4e} & =\frac{1}{2\pi\left(1-\kappa^{2}\right)}\\
J_{\mathrm{nem}} & =J_{4e}-\frac{1}{8\pi}\left[\frac{2}{1-\kappa^{2}}-\frac{1}{\kappa}\ln\left(\frac{1+\kappa}{1-\kappa}\right)\right]
\end{align}
which implies that $J_{4e}>J_{\mathrm{nem}}$, i.e. $r_{4e}>r_{\mathrm{nem}}$.
Now, $r$ is generally a decreasing function of temperature, since
$r\propto\xi^{-2}\rightarrow0$ at the bare superconducting transition
temperature (here $\xi$ is the superconducting correlation length).
Consequently, the $\kappa$ term in the action favors the charge-$4e$
instability over the nematic instability.

Note that this analysis is valid for the case of triangular or hexagonal
lattices. For trigonal lattices, there is an additional allowed term
in the susceptibility (\ref{chi_S2}) that depends on $q_{z}$ \citep{Venderbos_Fu_Hc2}.
Such a term generates the cubic invariant $\Phi_{+}^{3}+\Phi_{-}^{3}$
in the nematic free energy \citep{Hecker18}, with $\Phi_{\pm}=\Phi_{1}\pm\Phi_{2}$,
which favors the nematic instability over the charge-$4e$ instability.
Alternatively, this cubic term is expected to be generated from the
sixth-order term of the superconducting action discussed in the main
text.

\end{document}